\newif\ifAnon\Anonfalse

\newif\ifDraft\Draftfalse

\newcommand{\VariantOne}{Variant 1\xspace}
\newcommand{\VariantTwo}{Variant 2\xspace}

\documentclass[sigconf,nonacm]{acmart}

\renewcommand\footnotetextcopyrightpermission[1]{} 
\usepackage{tugraz_defaults}
\usepackage{enumitem}
\usepackage{subcaption}
\newcommand{\paragrabf}[1]{\paragraph{\textbf{#1}}}

\newcommand{\AttackName}{ZombieLoad\xspace}

\newcommand{\linuxlogo}{\includegraphics[height=0.9em]{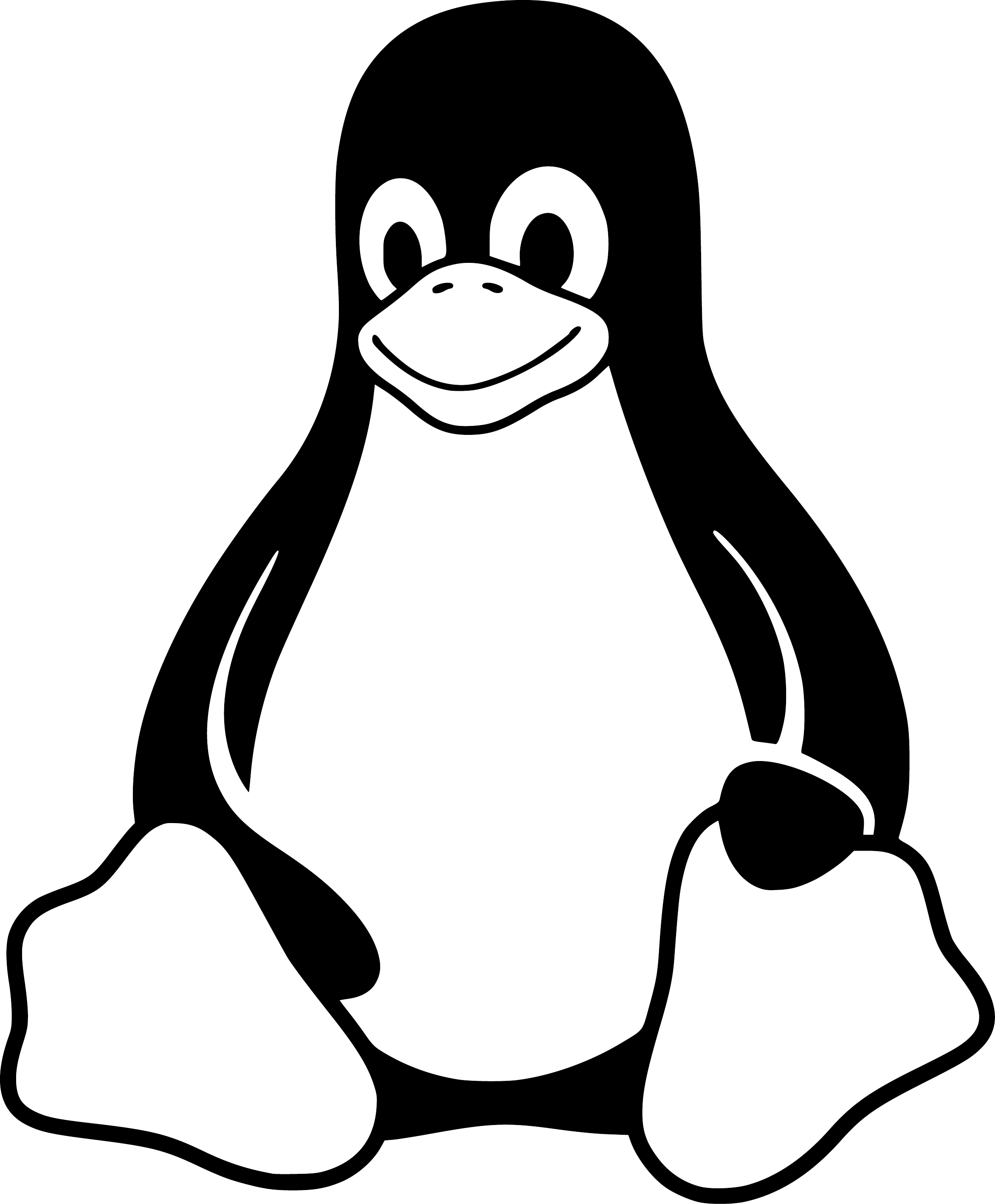}}
\newcommand{\winlogo}{\includegraphics[height=0.9em]{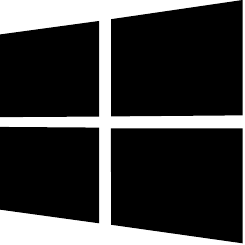}}

\begin{document}


\date{}
 
\title{\AttackName: Cross-Privilege-Boundary Data Sampling}

\author{Michael Schwarz}
\affiliation{\institution{Graz University of Technology}}
\email{michael.schwarz@iaik.tugraz.at}

\author{Moritz Lipp}
\affiliation{\institution{Graz University of Technology}}
\email{moritz.lipp@iaik.tugraz.at}

\author{Daniel Moghimi}
\affiliation{\institution{Worcester Polytechnic Institute}}
\email{amoghimi@wpi.edu}

\author{Jo Van Bulck}
\affiliation{\institution{imec-DistriNet, KU Leuven}}
\email{jo.vanbulck@cs.kuleuven.be}

\author{Julian Stecklina}
\affiliation{\institution{Cyberus Technology}}
\email{julian.stecklina@cyberus-technology.de}

\author{Thomas Prescher}
\affiliation{\institution{Cyberus Technology}}
\email{thomas.prescher@cyberus-technology.de}

\author{Daniel Gruss}
\affiliation{\institution{Graz University of Technology}}
\email{daniel.gruss@iaik.tugraz.at}

%
\renewcommand{\shortauthors}{Schwarz et al.}

\begin{abstract}
In early 2018, Meltdown first showed how to read arbitrary kernel memory from user space by exploiting side-effects from transient instructions.
While this attack has been mitigated through stronger isolation boundaries between user and kernel space, Meltdown inspired an entirely new class of fault-driven transient execution attacks.
Particularly, over the past year, Meltdown-type attacks have been extended to not only leak data from the L1 cache but also from various other microarchitectural structures, including the FPU register file and store buffer.

In this paper, we present the \AttackName attack which uncovers a novel Meltdown-type effect in the processor's previously unexplored fill-buffer logic.
Our analysis shows that faulting load instructions (\ie loads that have to be re-issued for either architectural or microarchitectural reasons) may transiently dereference unauthorized destinations previously brought into the fill buffer by the current or a sibling logical CPU.
Hence, we report data leakage of recently loaded stale values across logical cores.
We demonstrate \AttackName{}'s effectiveness in a multitude of practical attack scenarios across CPU privilege rings, OS processes, virtual machines, and SGX enclaves.
We discuss both short and long-term mitigation approaches and arrive at the conclusion that disabling hyperthreading is the only possible workaround to prevent this extremely powerful attack on current processors.
\end{abstract}

%
%


%

\maketitle

\newcommand{\inst}[1]{\texttt{#1}}

\section{Introduction}
In 2018, Meltdown~\cite{Lipp2018meltdown} was the first microarchitectural attack completely breaching the security boundary between the user and kernel space and, thus, allowed to leak arbitrary data. 
While Meltdown was fixed using a stronger isolation between user and kernel space, the underlying principle turned out to be an entire class of transient-execution attacks~\cite{Canella2019}. 
Over the past year, researchers have demonstrated that Meltdown-type attacks cannot only leak kernel data to user space, but also leak data across user processes, virtual machines, and SGX enclaves~\cite{Vanbulck2018,Weisse2018foreshadowNG}.
Furthermore, data cannot only be leaked from the L1 cache but also from other microarchitectural structures, including the register file~\cite{Stecklina2018LazyFP}, the line-fill buffer~\cite{Lipp2018meltdown,VanSchaik2019RIDL}, and, as shown in concurrent work, the store buffer~\cite{Genkin2019storebuffer}. 

Instead of executing the instruction stream in order, most modern processors can re-order instructions while maintaining architectural equivalence, creating the illusion of an in-order machine.
Instructions then may already have been executed when the CPU detects that a previous instruction raises an exception. 
Hence, such instructions following the faulting instruction (\ie transient instructions) are rolled back. 
While the rollback ensures that there are no architectural effects, side effects might remain in the microarchitectural state. 
Most Meltdown-type data leaks exploit overly aggressive performance optimizations around out-of-order execution.

For many years, the microarchitectural state was considered invisible to applications, and hence security considerations were often limited to the architectural state. 
Specifically, microarchitectural elements often do not distinguish between different applications or privilege levels~\cite{Jang2016,Pessl2016,Schwarz2017SGX,Lipp2018meltdown,Schwarz2018KeyDrown,Evtyushkin2018BranchScope,Canella2019}.  

In this paper, we show that, first, there still are unexplored microarchitectural buffers, and second, both architectural and microarchitectural faults can be exploited.
With our notion of \enquote{microarchitectural faults}, \ie faults that cause a memory request to be re-issued internally without ever becoming architecturally visible, we demonstrate that Meltdown-type attacks can also be triggered without raising an architectural exception such as a page fault.
Based on this, we demonstrate \AttackName, a novel, extremely powerful Meltdown-type attack targeting the fill buffer logic.

\AttackName exploits that load instructions which have to be re-issued internally, may first transiently compute on stale values belonging to previous memory operations from either the current or a sibling hyperthread.
Using established transient execution attack techniques, adversaries can recover the values of such \enquote{zombie load} operations.
Importantly, in contrast to all previously known transient execution attacks~\cite{Canella2019}, \AttackName reveals recent data values \emph{without} adhering to any explicit address-based selectors.
Hence, we consider \AttackName an instance of a novel type of \textit{microarchitectural data sampling} attacks.
We present microarchitectural data sampling as the missing link between traditional memory-based side-channels which correlate data adresses within a victim execution, and existing Meltdown-type transient execution attacks that can directly recover data values belonging to an explicit address.
In this paper, we combine primitives from traditional side-channel attacks with incidental data sampling in the time domain to construct extremely powerful attacks with targeted leakage in the address domain.
This not only opens up new attack avenues, but also re-enables attacks that were previously assumed to be mitigated. 

We demonstrate \AttackName's real-world implications in a multitude of practical attack scenarios that leak across processes, privilege boundaries, and even across logical CPU cores. 
Furthermore, we show that we can leak Intel SGX enclave secrets loaded from a sibling logical core.
We demonstrate that \AttackName attackers may extract sealing keys from Intel's architectural quoting enclave, ultimately breaking SGX's confidentiality and remote attestation guarantees. 
\AttackName is furthermore not limited to native code execution, but also works across virtualization boundaries. 
Hence, virtual machines can attack not only the hypervisor but also different virtual machines running on a sibling logical core. 
We conclude that disabling hyperthreading, in addition to flushing several microarchitectural states during context switches, is the only possible workaround to prevent this extremely powerful attack. 

\paragraph{Contributions.}
The main contributions of this work are:
\begin{enumerate}
  \item We present \AttackName, a powerful data sampling attack leaking data accessed on the same or sibling hyperthread.
  \item We combine incidental data sampling in the time domain with traditional side-channel primitives to construct a targeted information flow similar to regular Meltdown attacks.
  \item We demonstrate \AttackName in several real-world scenarios: cross-process, cross-VM, user-to-kernel, and SGX.
  \item We show that \AttackName breaks the security guarantees provided by Intel SGX.
  \item We are the first to do post-processing of the leaked data within the transient domain to eliminate noise.
\end{enumerate}

\paragraph{Outline.}
\cref{sec:background} provides background.
\cref{sec:attack-overview} provides an overview of \AttackName, and introduces a novel classification scheme for memory-based side-channel attacks. 
\cref{sec:attack-scenarios} describes attack scenarios and the respective attacker models. 
\cref{sec:building-blocks} introduces and evaluates the basic primitives required for mounting \AttackName.
\cref{sec:attacks} demonstrates \AttackName in real-world attack scenarios.
\cref{sec:countermeasures} discusses possible countermeasures. 
We conclude in \cref{sec:conclusion}.

\paragraph{Responsible Disclosure.}
We provided Intel with a PoC leaking uncacheable-typed memory locations from a concurrent hyperthread on March 28, 2018.
We clarified to Intel on May 30, 2018, that we attribute the source of this leakage to the LFB.
In our experiments, this works identically for Foreshadow (Meltdown-P), undermining the completeness of L1-flush-based mitigations.
This issue was acknowledged by Intel and tracked under CVE-2019-11091.
We responsibly disclosed the main attack presented in this paper to Intel on April 12, 2019. 
Intel verified and acknowledged our findings and assigned CVE-2018-12130 to this issue.
Both issues were part of an embargo ending on May 14, 2019.

\section{Background}
\label{sec:background}
In this section, we describe the background required for this paper.

\subsection{Transient Execution Attacks}

Today's high-performance processors typically implement an \textit{out-of-order execution} design, allowing the CPU to utilize different execution units in parallel.
The instruction stream is decoded \textit{in-order} into simpler micro-operations (\muops)~\cite{Fog2016} which can be executed as soon as the required operands are available.
A dedicated reorder buffer stores intermediate results and ensures that instruction results are committed to the architectural state in the order specified by the program's instruction stream.
Any fault that occurred during the execution of an instruction is handled at instruction retirement, leading to a pipeline flush which squashes any outstanding \muop results from the reorder buffer.

In addition, modern CPUs employ \textit{speculative execution} optimizations to avoid stalling the instruction pipeline until
a conditional branch is resolved.
The processor predicts the most likely outcome of the branch and continues execution along that direction.
If the branch is resolved and the prediction was correct, the speculative results retire in-order yielding a measurable performance improvement.
On the other hand, if the prediction was wrong, the pipeline is flushed, and any speculative results are squashed in the reorder buffer.
We refer to instructions that are executed speculatively or out-of-order but whose results are never architecturally committed as \textit{transient instructions}~\cite{Canella2019,Lipp2018meltdown,Vanbulck2018}.

While the results and the architectural effects of transient instructions are discarded, measurable microarchitectural side effects may remain and are not reverted.
Attacks that exploit these side effects to observe sensitive information are called \textit{transient execution attacks}~\cite{Lipp2018meltdown,Kocher2019spectre,Canella2019}.
Typically, these attacks utilize a cache-based covert channel to transmit the secret data observed transiently from the microarchitectural domain to an architectural state.
However, other covert channels can be utilized as well~\cite{Schwarz2018netspectre,Bhattacharyya2019smotherspectre}.
In line with a recent exhaustive survey~\cite{Canella2019}, we refer to attacks exploiting misprediction~\cite{Kocher2019spectre,Kiriansky2018speculative,Koruyeh2018spectre5,Maisuradze2018spectre5,Horn2018spectre4} as Spectre-type, whereas attacks exploiting transient execution after a CPU exception~\cite{Lipp2018meltdown,Vanbulck2018,Stecklina2018LazyFP,Weisse2018foreshadowNG,Kiriansky2018speculative,Canella2019} are classified as belonging to Meltdown-type.

\subsection{Memory Subsystem}\label{sec:background:loadbuffer}

The CPU architecture defines different instructions to load data from memory.
In this section, we give a high-level overview of how out-of-order CPUs handle memory loads.
However, as the actual implementation of the microarchitecture is usually not publicly documented, we rely on patents held by Intel to back up possible implementation details.

\paragrabf{Caches}
To improve the performance of memory accesses, CPUs contain small and fast internal caches that store frequently used data.
Caches are typically organized in multiple levels that are either private per core or shared amongst them.
Modern CPUs typically use $n$-way set-associative caches containing $n$ cache lines per set, each typically \SI{64}{\byte} wide.
Usually, modern Intel CPUs have a private first-level instruction (L1I) and data cache (L1D) and a unified L2 cache.
The last-level cache (LLC) is shared across all cores.

\paragrabf{Virtual Memory}
CPUs use virtual memory to provide memory isolation between processes.
Virtual addresses are translated to physical memory locations using multi-level translation tables. 
The translation table entries define the properties, \eg access control or memory type, of the referenced memory region.
The CPU contains the translation-look-aside buffer (TLB) consisting of additional caches to store address-translation information.

\paragrabf{Memory Order Buffer}
\muops that deal with memory operations are handled by dedicated execution units.
Typically, Intel CPUs contain 2 units responsible for loading data and one for storing data.
While the reorder buffer resolves register dependencies, out-of-order executed \muops can still have memory dependencies.
In an out-of-order CPU, the \textit{memory orde}r buffer (MOB), incorporating a \textit{load buffer} and a \textit{store buffer}, controls the dispatch of memory operations and tracks their progress to resolve memory dependencies.

\paragrabf{Data Loads}
For every dispatched load operation an entry is allocated in the load buffer and in the reorder buffer.
The allocated load-buffer entry holds information about the operation, \eg ordering constraints, the reorder buffer ID or the age of the most recent store.
To determine the physical address, the upper \SI{36}{\bit} of the linear address are translated by the memory management unit. 
Concurrently, the untranslated lower \SI{12}{\bit} are already used to index the cache set in the L1D~\cite{US5613083}.
If the address translation is cached in the TLB, the physical address is available immediately.
Otherwise, the page miss handler (PMH) is activated to perform a page-table walk to retrieve the address translation as well as the corresponding permission bits.
With the physical address, the tag and, thus, the way of the cache is determined.
If the requested data is in the L1D (cache hit), the load operation can be completed.

If data is not in the L1D, it needs to be served from higher levels of the cache or the main memory via the line-fill buffer (LFB).
The LFB serves as an interface to other caches and the main memory and keeps track of outstanding loads.
Memory accesses to uncacheable memory regions, and non-temporal moves all go through the LFB.
If a load corresponds to an entry of a previous load operation in the load buffer, the loads can be merged~\cite{US7346735,Abramson1996}.

On a fault, \eg a physical address is not available, the page-table walk will not immediately abort~\cite{US5613083}.
Still, an instruction in a pipelined implementation must undergo each stage regardless of whether a fault occurred or not~\cite{US5717882}, and is reissued in case of a fault.
Only at the retirement of the faulting \muop, the fault is handled, and the pipeline is flushed~\cite{US5613083,US5564111}.
If a fault occurs within a load operation, it is still marked as ``valid and completed'' in the MOB~\cite{US5717882}.

\subsection{Processor Extensions}\label{sec:background-extensions}

\paragrabf{Microcode}\label{sec:background:ucode}

Initially, all instructions were hardwired in the CPU core.
However, to support more complex instructions, \textit{microcode} allows implementing higher-level instructions using multiple hardware-level instructions.
Importantly, this allows processor vendors to support complex behavior and even extend or modify CPU behavior through microcode updates~\cite{Intel_vol3}.
Preferably, new architectural features are implemented as microcode extensions, \eg Intel SGX~\cite{US20120159184A1}.

While the execution units perform the fast-paths directly in hardware, more complex slow-path operations are typically performed by issuing a \textit{microcode assist} which points the sequencer to a predefined microcode routine~\cite{Costan2016}.
To do so, the execution unit associates an event code with the result of the faulting micro-op.
When the micro-op of the execution unit is committed, the event code causes the out-of-order scheduler to squash all in-flight micro-ops in the reorder buffer~\cite{Costan2016}.
The microcode sequencer uses the event code to read the micro-ops associated with the event in the microcode~\cite{US5625788A}.

\paragrabf{Intel TSX}\label{sec:background:tsx}

Intel TSX is an x86 instruction set extension to support hardware transactional memory~\cite{IntelTSX_CPP} which has been introduced with Intel Haswell CPUs.
With TSX, particular code regions are executed transactionally.
If the entire code regions completes successfully, memory operations within the transaction appear as an atomic commit to other logical processors.
If an issue occurs during the transaction, a transactional abort rolls back the execution to an architectural state before the transaction and, thereby, discarding all performed operations.
Transactional aborts can be caused by different issues: Typically, a conflicting memory operation occurs where another logical processor either reads from an address which has been modified within the transaction or writes to an address which is used within the transaction.
Further, the amount of read and written data within the transaction may not exceed the size of the LLC and L1 cache respectively for the transaction to succeed~\cite{Intel_vol3}.
In addition, some instructions or system event might cause the transaction to abort as well~\cite{IntelTSX_CPP}.

\paragrabf{Intel SGX}\label{sec:sgx}

With the Skylake microarchitecture, Intel introduced Software Guard Extension (SGX), an instruction-set extension for isolating trusted code~\cite{Intel_vol3}.
SGX executes trusted code inside so-called \textit{enclaves}, which are mapped in the virtual address space of a conventional host application process but are isolated from the rest of the system by the hardware itself.
The threat model of SGX assumes that the operating system and all other running applications could be compromised and, therefore, cannot be trusted.
Any attempt to access SGX enclave memory in non-enclave mode results in abort page semantics, \ie regardless of the current privilege level, reads return the dummy value \texttt{0xff} and writes are ignored~\cite{sgxdeveloperref}.
Furthermore, to protect against powerful physical attackers probing the memory bus, the SGX hardware transparently encrypts the memory region used by enclaves~\cite{Costan2016}.

A dedicated \texttt{eenter} instruction redirects control flow to an enclave entry point, whereas \texttt{eexit} transfers back to the untrusted host application.
Furthermore, in case of an interrupt or fault, SGX securely saves CPU registers inside the enclave's save state area (SSA) before vectoring to the untrusted operating system.
Next, the \texttt{eresume} instruction can be used to restore processor state from the SSA frame and continue a previously interrupted enclave.

SGX-capable processors feature cryptographic key derivation facilities through the \texttt{egetkey} instruction,
based on a CPU-level master secret and a secure measurement of the calling enclave's initial code and data.
Using this key, enclaves can securely \textit{seal} secrets for untrusted persistent storage, and establish secure communication channels with other enclaves residing on the same processor.
Furthermore, to enable remote attestation, Intel provides a trusted \textit{quoting enclave} which unseals an Intel-private key and generates an asymmetric signature over the local enclave identity report.

Over the past years, researchers have demonstrated various attacks to leak sensitive data from SGX enclaves, \eg through memory safety violations~\cite{Lee2017SGXROP}, race conditions~\cite{Weichbrodt2016}, or side-channels~\cite{Moghimi2017cachezoom,Schwarz2017SGX,Vanbulck2017pagetable,Vanbulck2018nemesis}.
More recently, SGX was also compromised by transient execution attacks~\cite{Vanbulck2018,Chen2018SGXpectre} which necessitated microcode updates and increased the processor's security version number (SVN).
All SGX key derivations and attestations include SVN to reflect the current microcode version, and hence security level.

\section{Attack Overview}\label{sec:attack-overview}

In this section, we provide an overview of \AttackName. 
We describe what can be observed using \AttackName and how that fits into the landscape of existing side-channel attacks. 
By that, we show that \AttackName is a novel category of side-channel attacks, which we refer to as \textit{data-sampling attacks}, opening a new research field. 

\subsection{Overview}
\AttackName is a transient-execution attack~\cite{Canella2019} which observes the values of memory loads on the current physical CPU. 
\AttackName exploits that the fill buffer is accessible by all logical CPUs of a physical CPU core and that it does not distinguish between processes or privilege levels. 

The load buffer acts as a queue for all memory loads from the memory subsystem. 
Whenever the CPU encounters a memory load during execution, it reserves an entry in the load buffer. 
If the load was not an L1 hit, it additionally requires a fill-buffer entry. 
When the requested data has been loaded, the memory subsystem frees the corresponding load- and fill-buffer entries, at which point the corresponding load instruction may retire.

However, we observed that under certain complex microarchitectural conditions (\eg a fault), where the load requires a microcode assist, it may first read stale values before being re-issued eventually.
As with any Meltdown-type attack, this opens up a transient execution window in which this value can be used for subsequent calculations before the execution is aborted and rolled back. 
Thus, an attacker can encode the leaked value into a microarchitectural element, such as the cache. 

In contrast to previous Meltdown-type attacks, however, it is not possible to select the value to leak based on an attacker-specified address. 
\AttackName simply leaks any value which is currently loaded by the physical CPU core. 
While this at first sounds like a massive limitation, we show that this opens a new field of side-channel attacks. 
We show that \AttackName is an even more powerful attack when combined with existing techniques known from traditional side-channel attacks. 

\subsection{Microarchitectural Root Cause}\label{sec:root-cause} 

For Meltdown, Foreshadow, and Fallout, the source of the leakage is apparent. 
Moreover, for these attacks, there are plausible explanations on what is going wrong in the microarchitecture, \ie what the root cause of the leakage is~\cite{Lipp2018meltdown,Vanbulck2018,Weisse2018foreshadowNG,Genkin2019storebuffer}. 
For \AttackName, however, this is not entirely clear. 

While we identified some necessary building blocks to observe the leakage (\cf \Cref{sec:building-blocks}), we can only provide a hypothesis on why the interaction of the building blocks leads to the observed leakage. 
As we could only observe data leakage on Intel CPUs, we assume that this is indeed an implementation issue (such as Meltdown) and not an issue with the underlying design (as with Spectre). 
For our hypothesis, we combined our observations with the nearly non-existent official documentation of the fill buffer~\cite{Intel_opt,Intel_vol3}. 
Ultimately, we could neither prove nor disprove our hypothesis, leaving the verification or falsification of our hypothesis to future work. 

\paragrabf{Stale-Entry Hypothesis.}
Every load is associated with an entry in the load buffer and potentially an entry in the fill buffer~\cite{Intel_opt}. 

When a load encounters a complex situation, such as a fault, it requires a microcode assist~\cite{Intel_vol3}. 
This microcode assist triggers a machine clear, which flushes the pipeline. 
On a pipeline flush, instructions which are already in flight still finish execution~\cite{Hennessy2017}. 

As this has to be as fast as possible to not incur additional delays, we expect that fill-buffer entries are optimistically matched as long as parts of the physical address match. 
Thus, the load continues with a wrong fill-buffer entry, which was valid for a previous load. 
This leads to a use-after-free vulnerability~\cite{Gruss2018useafterfree} in the hardware. 
Intel documents the fill buffer as being competitively shared among hyperthreads~\cite{Intel_vol3}, giving both logical cores access to the entire fill buffer (\cf \Cref{appendix:fill-buffer-size}). 
Consequently, the stale fill-buffer entry can also be from a previous load of the sibling logical core. 
As a result, the load instruction loads valid data from a previous load. 

\paragrabf{Leakage Source.}
We devised 2 experiments to reduce the number of possible sources of the leaked data. 

In our first experiment, we marked a page as \enquote{uncacheable} via the page-table entry and flushed the page from the cache. 
As a result, every memory load from the page circumvents all cache levels and directly travels from the main memory to the fill buffer~\cite{Intel_vol3}. 
We then write the secret onto the uncacheable memory page to ensure that there is no copy of the data in the cache. 
When loading data from the uncacheable memory page, we can see leakage, but the leakage rate is only in the order of bytes per second, \eg \SI{5.91}{\byte/\second} ($\sigma_{\bar{x}}=0.18$, $n=100$) on an i7-8650U. 
We can attribute this leakage to the fill buffer. 
This was also exploited in concurrent work~\cite{VanSchaik2019RIDL}. 
Our hypothesis is further backed by the \texttt{MEM\_LOAD\_RETIRED.FB\_HIT} performance counter, which shows multiple thousand line-fill-buffer hits (\SI{117330}{FB\_HIT/\second} ($\sigma_{\bar{x}}=511.57$, $n=100$)). 

Intel claims that the leakage is entirely from the fill buffer. 
However, our second experiment shows that the line-fill buffer might not be the only source of the leakage. 
We rely on Intel TSX to ensure that memory accesses do not reach the line-fill buffer as follows. 
Inside a transaction, we first write the secret value to a memory location which was previously initialized with a different value. 
The write inside the transaction ensures that the address is in the \textit{write set} of the transaction and thus in L1~\cite{Intel_opt,Schwarz2018DF}. 
Evicting data from the write set from the cache leads to a transactional abort~\cite{Intel_opt}.  
Hence, any subsequent memory access to the data from the write set ensures that it is served from the L1, and therefore, no request to the line-fill buffer is sent~\cite{Intel_vol3}. 
In this experiment, we see a much higher rate of leakage which is in the order of kilobytes per second. 
More importantly, we only see the value written inside the TSX transaction and not the value that was at the memory location before starting the transaction. 
Our hypothesis that the line-fill buffer is not the only source of the leakage is further backed by observing performance counters. 
The \texttt{MEM\_LOAD\_RETIRED.FB\_HIT} and \texttt{MEM\_LOAD\_RETIRED.L1\_MISS} performance counters, do not increase significantly. 
In contrast, the \texttt{MEM\_LOAD\_RETIRED.L1\_HIT} performance counter shows multiple thousand L1 hits. 

While accessing the data to leak on the victim core, we monitored the \texttt{MEM\_LOAD\_RETIRED.FB\_HIT} performance counter on the attacker core for \SI{10}{\second}.
If the address was cached, we measured a Pearson correlation of $r_p=0.02$ ($n=100$) between the correct recoveries and line-fill buffer hits, indicating no association.
However, while continuously flushing the data on the victim core, ensuring that a subsequent access must go through the LFB, we measure a strong correlation of $r_p=0.86$ ($n=100$).
This result indicates that the line-fill buffer is not the only source of leakage.
However, a different explanation might be that the performance counters are not reliable in such corner cases. 
Future work has to investigate whether other microarchitectural elements, \eg the load buffer, are also involved in the observed data leakage. 

\subsection{Classification}

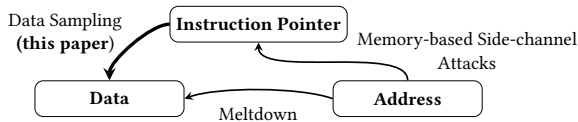
\begin{figure}[t]
 \centering
 \resizebox{\hsize}{!}{
    \begin{tikzpicture}[node distance=2.5cm]
\tikzstyle{small}  = [rectangle, rounded corners, minimum width=2.5cm, minimum height=.6cm,text centered, draw=black, fill=white]
\tikzstyle{arrow}  = [thick,->,>=stealth,in=180,out=0,looseness=0.6]

\usetikzlibrary{shapes.geometric, arrows, patterns}

\node (rip) [small] {\textbf{Instruction Pointer}};
\node (addr) [small,right of=rip,below of=rip,yshift=1.25cm] {\textbf{Address}};
\node (data) [small,left of=rip,below of=rip,yshift=1.25cm] {\textbf{Data}};

\draw (addr.west) edge[thick,->,>=stealth,in=20,out=160,looseness=0.6] (data.east) node[xshift=-1.25cm,yshift=-0.25cm] {Meltdown};

\draw (addr.north) edge[thick,->,>=stealth,in=270,out=90,looseness=0.6] (rip.south) node[xshift=1cm,yshift=0.5cm] {\parbox{4cm}{\centering Memory-based Side-channel Attacks}};

\draw (rip.west) edge[ultra thick,->,>=stealth,in=90,out=180,looseness=0.6] (data.north) node[xshift=-1.75cm,yshift=-0.15cm] {\parbox{3cm}{\centering Data Sampling\\\textbf{(this paper})}};

\end{tikzpicture}
 }
 \caption{The 3 properties of a memory operation: instruction pointer of the program, target address, and data value. 
 So far, there are techniques to infer the instruction pointer from target address and the data value from the address. 
 With \AttackName, we show the first instance of an attack which infers the data value from the instruction pointer.}
 \label{fig:triangle}
\end{figure}

In this section, we introduce a way to classify memory-based side-channel and transient-execution attacks. 
For all these attacks, we assume a target program which executes a memory operation at a certain \textit{address} with a specific data \textit{value} at the program's current \textit{instruction pointer}.  
\Cref{fig:triangle} illustrates these three properties as the corner of a triangle, and techniques which let an attacker infer one of the properties based on one or both of the other properties. 

Traditional memory-based side-channel attacks allow an attacker to observe the location of memory accesses. 
The granularity of the location observation depends on the spatial accuracy of the used side channel. 
Most common memory-based side-channel attacks~\cite{Percival2005,Yarom2014,Gruss2016Flush,Gruss2016Prefetch,Gruss2015Template,Pessl2016,Xu2015controlled,Vanbulck2017pagetable,Jang2016,Gras2018TLB} have a granularity between one cache line~\cite{Yarom2014,Gruss2016Flush,Gruss2016Prefetch,Gruss2015Template} \ie usually \SI{64}{\byte}, and one page~\cite{Jang2016,Gras2018TLB,Vanbulck2017pagetable,Xu2015controlled}, \ie usually \SI{4}{\kilo\byte}. 
These side channels establish a connection between the time domain and the space domain. 
The time domain can either be the wall time or also commonly the execution time of the program which correlates with the instruction pointer. 
These classic side channels provide means of connecting the address of a memory access to a set of possible instruction pointers, which then allows reconstructing the program flow. 
Thus, side-channel resistant applications have to avoid secret-dependent memory access to not leak secrets to a side-channel attacker.

\begin{figure}[t]
 \centering
 \resizebox{\hsize}{!}{
    \resizebox{\hsize}{!}{
\begin{tikzpicture}[xscale=.75,yscale=1.25]
\usetikzlibrary{patterns}
\begin{scope}[shift={(0,-3+0.125*3)}]
\draw[densely dotted,fill=black!0] (0,3.25) rectangle +(8,0.25);
\draw[densely dotted,fill=black!0] (0.5,3) rectangle +(7.5,0.25);
\draw[densely dotted,fill=black!0] (8,3) rectangle +(2,0.5);
\draw[postaction={draw,pattern=north east lines, pattern color=black!40}, fill=black!20] (10,3) rectangle +(2,0.5);
\node at (7.75,3.375) {\scriptsize 12};
\node[text centered] at (4,3.375) {\scriptsize Physical};
\node at (7.75,3.125) {\scriptsize 12};
\node[text centered] at (4,3.125) {\scriptsize Virtual};
\node[align=right,text width=1.6cm] at (-1.5,3.25) {\AttackName};
\node at (0.25,3.375) {\scriptsize 51};
\node at (0.75,3.125) {\scriptsize 47};
\node at (8.25,3.25) {\scriptsize 11};
\node at (9.75,3.25) {\scriptsize 6};
\node at (10.25,3.25) {\scriptsize 5};
\node at (11.75,3.25) {\scriptsize 0};
\end{scope}

\begin{scope}[shift={(0,-2.25+0.125*2)}]
\draw[densely dotted,fill=black!0] (0,3.25) rectangle +(8,0.25);
\draw[densely dotted,fill=black!0] (0.5,3) rectangle +(7.5,0.25);
\draw[postaction={draw,pattern=north east lines, pattern color=black!30}, fill=black!20] (8,3) rectangle +(4,0.5);
\node at (7.75,3.375) {\scriptsize 12};
\node[text centered] at (4,3.375) {\scriptsize Physical};
\node at (7.75,3.125) {\scriptsize 12};
\node[text centered] at (4,3.125) {\scriptsize Virtual};
\node[align=right,text width=1.6cm] at (-1.5,3.25) {Fallout};
\node at (0.25,3.375) {\scriptsize 51};
\node at (0.75,3.125) {\scriptsize 47};
\node at (8.25,3.25) {\scriptsize 11};
\node at (11.75,3.25) {\scriptsize 0};
\end{scope}

\begin{scope}[shift={(0,-1.5+0.125)}]
\draw[postaction={draw,pattern=north east lines, pattern color=black!30}, fill=black!20] (0,3.25) rectangle +(8,0.25);
\draw[densely dotted,fill=black!0] (0.5,3) rectangle +(7.5,0.25);
\draw[postaction={draw,pattern=north east lines, pattern color=black!30}, fill=black!20] (8,3) rectangle +(4,0.5);
\node at (7.75,3.375) {\scriptsize 12};
\node[text centered] at (4,3.375) {\scriptsize Physical};
\node at (7.75,3.125) {\scriptsize 12};
\node[text centered] at (4,3.125) {\scriptsize Virtual};
\node[align=right,text width=1.6cm] at (-1.5,3.25) {Foreshadow};
\node at (0.25,3.375) {\scriptsize 51};
\node at (0.75,3.125) {\scriptsize 47};
\node at (8.25,3.25) {\scriptsize 11};
\node at (11.75,3.25) {\scriptsize 0};
\end{scope}

\begin{scope}[shift={(0,-0.75)}]
\draw[densely dotted,fill=black!0] (0,3.25) rectangle +(8,0.25);
\draw[postaction={draw,pattern=north east lines, pattern color=black!30}, fill=black!20] (0.5,3) rectangle +(8,0.25);
\draw[postaction={draw,pattern=north east lines, pattern color=black!30}, fill=black!20] (8,3) rectangle +(4,0.5);
\node at (7.75,3.375) {\scriptsize 12};
\node[text centered] at (4,3.375) {\scriptsize Physical};
\node at (7.75,3.125) {\scriptsize 12};
\node[text centered] at (4,3.125) {\scriptsize Virtual};
\node[align=right,text width=1.6cm] at (-1.5,3.25) {Meltdown};
\node at (0.25,3.375) {\scriptsize 51};
\node at (0.75,3.125) {\scriptsize 47};
\node at (8.25,3.25) {\scriptsize 11};
\node at (11.75,3.25) {\scriptsize 0};
\end{scope}

\node at (4,3) {Page Number};
\node at (10,3) {Page Offset};
\end{tikzpicture}
}
 }
 \caption{Meltdown-type attacks provide a varying degree of target control (gray hatched), from full virtual addresses in the case of Meltdown to nearly no control for \AttackName.}
 \label{fig:control}
\end{figure}
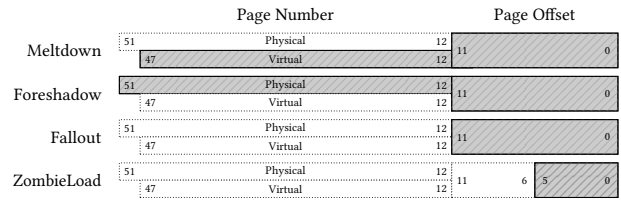

Since early 2018, with transient execution attacks~\cite{Canella2019} such as Meltdown~\cite{Lipp2018meltdown} and Spectre~\cite{Kocher2019spectre}, there is a second type of attacks which allow an attacker to observe the value stored at a memory address.
Meltdown provided the most control over target address. 
With Meltdown, the full virtual address of the target data is provided, and the corresponding data value stored at this address is leaked. 
The success rate depends on the location of the data, \ie whether it is in the cache or main memory. 
However, the only constraint for Meltdown is that the data is addressable using a virtual address~\cite{Lipp2018meltdown}. 
Other Meltdown-type attacks~\cite{Vanbulck2018,Genkin2019storebuffer} also connect addresses to data values. 
However, they often impose additional constraints, such as that the data has to be cached in L1~\cite{Vanbulck2018,Weisse2018foreshadowNG}, the physical address has to be known~\cite{Weisse2018foreshadowNG}, or that an attacker can choose only parts of the target address~\cite{Genkin2019storebuffer}. 

\Cref{fig:control} illustrates which parts of the virtual and physical address an attacker can choose to target data values to leak. 
For Meltdown, the virtual address is sufficient to target data in the same address space~\cite{Lipp2018meltdown}.
Foreshadow already requires knowledge of the physical address and the least-significant 12 bits of the virtual address to target any data in the L1, not limited to the own address space~\cite{Vanbulck2018,Weisse2018foreshadowNG}. 
When leaking the last writes from the store buffer, an attacker is already limited in choosing which value to leak. 
It is only possible to filter stores based on the least-significant 12 bits of the virtual address, a more targeted leakage is not possible~\cite{Genkin2019storebuffer}. 

Zombie loads provide no control over the leaked address to an attacker. 
The only possible target selection is the byte index inside the loaded data, which can be seen as an address with up to 6-bit in case an entire cache line is loaded.
Hence, we do not count \AttackName as an attack which leaks data values based on the address. 
Instead, from the viewpoint of the target control, \AttackName is more similar to traditional memory-based side-channel attacks. 
With \AttackName, an attacker observes the data value of a memory access. 
Thus, this side channel establishes a connection between the time domain and the data value. 
Again, the time domain correlates with the instruction pointer of the target address.
\AttackName is the first instance of a class of attacks which connects the \textit{instruction pointer} with the \textit{data value} of a memory access. 
We refer to such attacks as \textit{data sampling attacks}. 
Essentially, this new class of data sampling attacks is capable of breaking side-channel resistant applications, such as constant-time cryptographic algorithms~\cite{Gueron2012aesni}. 

Following the classification scheme from Canella~\etal\cite{Canella2019},
\AttackName is a Meltdown-type transient execution attack, and we propose \textit{Meltdown-MCA} as the generic name.
This reflects that the (microarchitectural) fault type being exploited by \AttackName is the microcode assist (MCA, explained further).

\section{Attack Scenarios \& Attacker Model}\label{sec:attack-scenarios} 

Following most side-channel attacks, we assume the attacker can execute unprivileged native code on the target machine. 
Thus, we assume a trusted operating system if not stated otherwise. 
This relatively weak attacker model is sufficient to mount \AttackName. 
However, we also show that the increased attacker capabilities offered in certain scenarios, \eg SGX and hypervisor attacks, may further amplify the leakage while remaining within the threat model of the respective scenario.  

At the hardware level, we assume a ubiquitous Intel CPU with simultaneous multithreading (SMT, also known as hyperthreading) enabled.
Crucially, we do not rely on existing vulnerabilities, such as Meltdown~\cite{Lipp2018meltdown}, Foreshadow~\cite{Vanbulck2018,Weisse2018foreshadowNG}, or Fallout~\cite{Genkin2019storebuffer}.

\paragrabf{User-Space Leakage}
In the cross-process user-space scenario, an unprivileged attacker leaks values loaded by another concurrently running user-space application.
We consider such a cross-process scenario most dangerous for end users, who are not commonly using Intel SGX nor virtual machines.
Moreover, many secrets are likely to be found in user-space applications such as browsers or password managers. 

The attacker can execute unprivileged code and is co-located with the victim on the same physical but a different logical CPU core. 
This is a typical case for hyperthreading, where both attacker and victim run on one hyperthread of the same CPU. 

\paragrabf{Kernel Leakage}
In addition to leakage across user-space applications, \AttackName can also leak across the privilege boundary between user and kernel space.
We demonstrate that the value of loads executed in kernel space is leaked to an unprivileged attacker, executing either on the same or a sibling logical core. 

In this scenario, the unprivileged attacker performs a system call to the kernel, running on the same logical core. 
Importantly, we found that kernel load leakage may even survive the switch back from the kernel to user space. 
Hyperthreading is hence \textit{not} a strict requirement for this scenario.

\paragrabf{Intel SGX Leakage}
In addition to leaking values loaded by the kernel, \AttackName can observe loads executed inside an Intel SGX enclave.
In this scenario, the attacker is executing on a sibling logical core, co-located with the victim enclave on the same physical core.
We demonstrate that \AttackName can leak secrets loaded during the enclave's execution from a concurrent logical core, but we did not observe leakage on the \emph{same} logical core after exiting the enclave synchronously (\texttt{eexit}) or asynchronously (on interrupt).

While in the aftermath of the Foreshadow~\cite{Vanbulck2018} attack, current SGX attestations indicate whether hyperthreading has been enabled at boot time, Intel's official security advisory~\cite{IntelL1TF} merely suggests that a remote verifier might reject attestations from a hyperthreading-enabled system \enquote{if it deems the risk of potential attacks from the sibling logical processor as not acceptable}.
Hence, machines with up-to-date patched microcode may still run with hyperthreading enabled.

Within the SGX threat model, we can leverage the attacker's first rate control over the untrusted operating system.
An attacker can, for instance, modify page table entries~\cite{Vanbulck2017pagetable}, or precisely execute the victim enclave at most one instruction at a time~\cite{VanBulck2017sgx}.

\paragrabf{Virtual Machine Leakage}
With \AttackName, it is possible to leak loaded values across virtual-machine boundaries.
In this scenario, an attacker running inside a virtual machine can leak values from a different virtual machine co-located on the same physical but different logical core.
Thus, an attacker can leak values loaded from a virtual machine running on the sibling logical core. 

As the attacker is running inside an untrusted virtual machine, the attacker is not restricted to unprivileged code execution.
Thus, the attacker can, for instance, modify guest-page-table entries. 

\paragrabf{Hypervisor Leakage}
In the hypervisor scenario, an attacker running inside a virtual machine utilizes \AttackName to leak the value of loads executed by the hypervisor. 

As the attacker is running inside an untrusted virtual machine, the attacker is not restricted to unprivileged code execution.

\section{Building Blocks}\label{sec:building-blocks}

In this section, we describe the building blocks for the attack. 

\subsection{Zombie Loads}\label{sec:faultingloads}

\begin{table}[t]
	\setlength{\aboverulesep}{0pt}
	\setlength{\belowrulesep}{0pt}
    \newcommand{\na}{\lineh} 
  \caption{Overview of different variants to induce zombie loads in different scenarios.}
  \label{tab:scenarios}
\begin{center}
\adjustbox{max width=\columnwidth}{\small{
		\setlength\tabcolsep{1.5pt}
        \begin{tabular}{r cc p{0.25cm} cc}
      \multirow{2}{*}{\diagbox{\textbf{Scenario}}{{\textbf{Variant}}}} & \multicolumn{2}{c}{\textbf{1}} & \quad & \multicolumn{2}{c}{\textbf{2}}   \\
      & \winlogo & \linuxlogo & & \winlogo & \linuxlogo  \\
			\toprule \vspace{-0.3cm}\\
		Unprivileged Attacker \quad	& \circlet 	& \circletfillhl	& & \circletfill 	& \circlet \\
		Privileged Attacker (root) \quad	& \circletfill 	& \circletfill			& & \circletfill 	& \circletfill 	 \\
			\bottomrule
		\end{tabular}
	}}
\end{center}
\footnotesize{
    Symbols indicate whether a variant can be used in the corresponding attack scenario
    (\circletfill), can be used depending on the hardware configuration as discussed in \cref{sec:faultingloads} (\circletfillhl), or cannot be used (\circlet).}
\end{table}

The main primitive for mounting \AttackName is a load which triggers a microcode assist, resulting in a transient load containing wrong data. 
We refer to such a load as a \textit{zombie load}. 
Zombie loads are loads which either architecturally or microarchitecturally fault and thus cannot complete, requiring a re-issue of the load at a later point. 
We identified multiple different scenarios to create such zombie loads required for a successful attack.
All variants have in common that they abuse the \texttt{clflush} instruction to reliably create the conditions required for leaking from a wrong destination (\cf \Cref{sec:root-cause}). 
In this section, we describe 2 different variants that can be used to leak data (\cf \Cref{sec:block-leakage}) depending on the adverary's capabilities.
\Cref{tab:scenarios} overviews which variant is applicable in which scenario, depending on the operating system and underlying hardware configuration.

\paragrabf{\VariantOne: Kernel Mapping.}

\begin{figure}
 \centering
 \resizebox{\hsize}{!}{
 \begin{tikzpicture}

\draw[dotted] (-1,3.5) -- (0,3) -- (0,0.5) -- (-1,1) -- (-1,3.5);
\draw[dotted] (-1,3.1) -- (0,2.6);
\draw[fill=black!2,dotted] (3,3.5) -- (2,3) -- (2,1.75) -- (3,2.25) -- (3,3.5);
\draw[dotted] (3,3.1) -- (2,2.6);

\draw (0,0.5) rectangle +(2,2.5) node[pos=.5,yshift=-0.65cm] {\parbox{2cm}{\centering Page \textit{p}\\2\,MB}};
\draw[fill=black!10] (0,1.75) rectangle +(2,1.25);

\draw[fill=black!10] (3,2.25) rectangle +(2,1.25) node[pos=.5,yshift=-0.25cm] {\parbox{2.5cm}{\centering User mapping\\\textit{v}}};
\node at (4,2) {4\,KB};
\node at (-2,0.75) {2\,MB};

\draw (-3,1) rectangle +(2,2.5) node[pos=.5,yshift=-0.1cm] {\parbox{2.5cm}{\centering Kernel\\address\\\textit{k}}};

\node[left] at (2.9,1.8) {\small 4\,KB};
\node[left] at (2.9,0.6) {\small 2\,MB};

\draw (-1,3.3) edge[->,>=stealth,out=-20,in=170,thick] (0,2.82);
\draw (3,3.3) edge[->,>=stealth,out=200,in=10,thick] (2,2.82);

\draw[fill=yellow] (0,2.6) rectangle +(2,0.4) node[pos=.5] {cache line};
\draw[fill=green!20] (3,3.1) rectangle +(2,0.4) node[pos=.5] {flush};
\draw[fill=red!20] (-3,3.1) rectangle +(2,0.4) node[pos=.5] {faulting load};

\end{tikzpicture}
 }
 \caption{\VariantOne: Using huge kernel pages for \AttackName.
 Page \textit{p} is mapped using a user-accessible address (\textit{v}) and a kernel-space huge page (\textit{k}).
 Flushing \textit{v} and then reading from \textit{k} using Meltdown leaks values from the fill buffer. }
 \label{fig:hugepage-setup}
\end{figure}
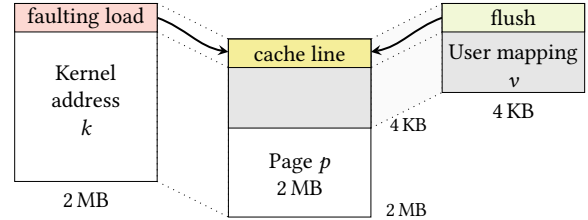

The first variant is a \AttackName setup which does not rely on any specific CPU feature.
We require a kernel virtual address $k$, \ie an address where the user-accessible bit is \textit{not} set in the page-table entry. 
In practice, the kernel is usually mapped with huge pages (\ie \SI{2}{\mega\byte} pages). 
Thus $k$ refers to a \SI{2}{\mega\byte} physical page $p$. 
Note that although we use such huge pages for our experiments, it is not strictly required, as the setup also works with \SI{4}{\kilo\byte} pages. 
We also require the user to have read access to the content of the physical page through a different virtual address $v$. 

\Cref{fig:hugepage-setup} illustrates such a setup.
In this setup, accessing the page $p$ via the user-accessible virtual address $v$ provides an architecturally valid way to access the contents of the page. 
Accessing the same page via the kernel address $k$ results in a zombie load similar to Meltdown~\cite{Lipp2018meltdown} requiring a microcode assist. 
Note that while there are other ways to construct an inaccessible address $k$, \eg by clearing the present bit~\cite{Vanbulck2018}, we were only able to exploit zombie loads originating from kernel mappings. 

To create precisely the scenario depicted in \Cref{fig:hugepage-setup}, we allocate a page \textit{p} in the user space with the virtual address \textit{v}.
Note that \textit{p} is a regular \SI{4}{\kilo\byte} page which is accessible through the virtual address \textit{v}.
We retrieve its physical address through \texttt{/proc/pagemap}, or alternatively using a side channel~\cite{Gruss2016Prefetch,Islam2019spoiler}. 
Using the physical address and the base address of the direct-physical map, we get an inaccessible kernel address \textit{k} which maps to the allocated page \textit{p}.
If the operating system does not use stronger kernel isolation~\cite{Gruss2017Kaslr}, \eg KPTI~\cite{LWN_kpti}, the direct-physical map in the kernel is mapped in the user space and uses huge pages which are marked as not user accessible. 
In the case of a privileged attacker, \eg when attacking a hypervisor or SGX enclave, an attacker can easily create such pages if they do not exist.

\paragrabf{\VariantTwo: Microcode-Assisted Page-Table Walk.}
A variant similar to \VariantOne is to trigger a microcode-assisted page-table walk. 
If a page-table walk requires an update to the access or dirty bit in the page-table entry, it falls back to a microcode assist~\cite{Costan2016}. 

In this setup, we require one physical page $p$ which has 2 user-accessible virtual addresses, $v$ and $v_2$. 
This can be easily achieved by using a shared-memory segment or memory-mapped file, which is mapped twice in the application.
The virtual address $v$ can be used to access the contents of $p$ architecturally. 
For $v_2$, we have to clear the accessed bit in the page-table entry. 
On Linux, this is not possible in the case of an unprivileged attacker, and can thus only be used in attacks where we assume a privileged attacker (\cf \Cref{sec:attack-scenarios}). 
However, we experimentally verified that Windows 10 (1803 build 17134.706) periodically clears the accessed bits. 
We assume that the page-replacement algorithm is responsible for this. 
Thus, this variant enables the attack on Windows for unprivileged attackers.

When accessing the page through the virtual address $v_2$, the accessed bit of the page-table entry has to be set. 
This, however, cannot be done by the page-miss handler~\cite{Costan2016}. 
Instead, microarchitecturally, the load faults, and a micro-code assist is triggered which repeats the page-table walk and sets the accessed bit~\cite{Costan2016}.

If the access to $v_2$ is done transiently, \ie behind a misspeculated branch or after an exception, the accessed bit cannot be set architecturally. 
Thus, the leakage is not only exploitable once but instead for every access.

\subsection{Data Leakage}\label{sec:block-leakage}

To leak data with the setup described in \Cref{sec:faultingloads}, we constantly flush the first cache line of \textit{p} through the virtual address \textit{v}. 
We achieve this by executing the unprivileged \texttt{clflush} instruction (or \texttt{clflushopt} instruction if available) on the user-accessible virtual address \textit{v}. 
For \VariantOne, we leverage Meltdown to read from the kernel address \textit{k} which maps to the cache line flushed before.
As with Meltdown-US~\cite{Lipp2018meltdown}, various methods of preventing an architectural exception can be used. 
We verified that \AttackName with \VariantOne works with exception prevention (\ie speculative execution), handling (\ie a custom signal handler), and suppression (\ie Intel TSX). 

For \VariantTwo, we transiently, \ie behind a mispredicted branch, read from the address $v_2$. 

Counterintuitively, the resulting values leaked for all variants are not coming from page \textit{p}. 
Instead, we get access to data which is currently loaded on the current or sibling logical CPU core. 
Thus, it appears that we reuse fill-buffer entries, and leak the data which the entries references. 
For \VariantOne and \VariantTwo, this allowed us to access all bytes from the cache line that the fill-buffer entry references. 

\subsection{Data Sampling}

Independent of the setup for \AttackName, we cannot directly control the address of the data to leak. 
Both the virtual addresses \textit{k} and \textit{v}, as well as the physical address of \textit{p} is arbitrary and does not correlate with the leaked data. 
In any case, we simply get the value referenced by one fill-buffer entry which we cannot specify. 

However, there is at least control within the fill-buffer entry, \ie we can target specific bytes \textit{within} the \SI{64}{\byte} fill-buffer entry. 
The least-significant 6 bits of the virtual address \textit{v} refer to the byte within the fill-buffer entry. 
Hence, we can target a single byte at a specific position from the fill-buffer entry. 
While at first, this does not sound powerful, it allows leaking sensitive information, such as AES keys, byte-by-byte as shown in \Cref{sec:attack-aes}. 

As described in \Cref{sec:attack-scenarios}, the leakage is not limited to the own process. 
With \AttackName, we observe values from all processes running on the same as well as on the sibling logical CPU core. 
Furthermore, we also observe leakage across privilege boundaries, \ie from the kernel, hypervisor, and Intel SGX enclaves. 
Thus, \AttackName allows sampling of all data which is loaded by any application on the current physical CPU core.

\subsection{Performance Evaluation}\label{sec:evaluation}
In this section, we evaluate \AttackName and the performance of our proof-of-concept implementations\footnote{\url{https://github.com/IAIK/ZombieLoad}}. 

\paragrabf{Environment}

\begin{table}[t]
\caption{Tested environments.}
\label{tab:environments}
\adjustbox{max width=\hsize}{
 \begin{tabular}{lllcc}
   \multicolumn{3}{c}{~} & 
   \multicolumn{2}{c}{\textbf{Variant}} \\
   \textbf{Setup} & \textbf{CPU} & \textbf{$\mu$-arch.} & \textbf{1} & \textbf{2}  \\ \hline
  Lab & Core i7-3630QM & Ivy Bridge & \cmark & \cmark \\
  Lab & Core i7-6700K & Skylake-S & \cmark & \cmark \\
  Lab & Core i5-7300U & Kaby Lake & \cmark & \cmark \\
  Lab & Core i7-7700 & Kaby Lake & \cmark & \cmark \\
  Lab & Core i7-8650U & Kaby Lake-R & \cmark & \cmark  \\
  Lab & Core i7-8565U & Whiskey Lake & \xmark & \xmark   \\
  Lab & Core i7-8700K & Coffee Lake-S & \cmark & \cmark  \\
  Lab & Core i9-9900K & Coffee Lake-R & \xmark & \xmark \\
  Lab & Xeon E5-1630 v4 & Broadwell-EP & \cmark & \cmark  \\
  Cloud & Xeon E5-2670 & Sandy Bridge-EP & \cmark & \cmark \\
  Cloud & Xeon Gold 5120 & Skylake-SP & \cmark & \cmark \\
  Cloud & Xeon Platinum 8175M & Skylake-SP & \cmark & \cmark  \\
  Cloud & Xeon Gold 5218 & Cascade Lake-SP & \xmark & \xmark  \\
  \hline
 \end{tabular}
 }
\end{table}

We evaluated the different variants of \AttackName, described in \Cref{sec:faultingloads}, on different environments listed in \Cref{tab:environments}.
The tested CPUs range from Sandy Bridge (released 2012) to Cascade Lake (released 2019).
We were able to mount \VariantOne and \VariantTwo on different microarchitectures except for Whiskey Lake, Coffee Lake-R, and Cascade Lake-SP.

\paragrabf{Performance}

To evaluate the performance of each variant, we performed the following experiment on an i7-8650U.
While reading a specific value on one logical core, we performed each variant of \AttackName on the sibling logical core for \SI{10}{\second}, recording the number of successful and unsuccessful recoveries.
For \VariantOne using TSX to suppress the exception, we achieve an average transmission rate of \SI{5.30}{\kilo\byte/\second} ($\sigma_{\bar{x}}=\SIx{0.076}$, $n=1000$) and a true positive rate of \SI{85.74}{\percent} ($\sigma_{\bar{x}}=0.0046$, $n=1000$).
With \VariantTwo in combination with signal handling, we achieved an average transmission rate of \SI{0.08}{\kilo\byte/\second} ($\sigma_{\bar{x}}=\SIx{0.002}$, $n=1000$) and a true positive rate of \SI{52.7}{\percent} ($\sigma_{\bar{x}}=0.0062$, $n=1000$).
\VariantTwo in combination with TSX, achieves an average transmission rate of \SI{7.73}{\kilo\byte/\second} ($\sigma_{\bar{x}}=\SIx{0.21}$, $n=1000$) and a true positive rate of \SI{76.28}{\percent} ($\sigma_{\bar{x}}=0.0055$, $n=1000$).

\section{Case Study Attacks}\label{sec:attacks} 
In this section, we present 5 attacks using \AttackName in real-world scenarios.

\subsection{AES-NI Key Leakage}\label{sec:attack-aes}

To demonstrate that data sampling is a powerful side channel, we extract an AES-128 key. 
The victim application uses AES-NI, which is resistant against timing and cache-based side-channel attacks~\cite{Gueron2012aesni}. 

However, even with the hardware-assisted AES-NI, the key has to be loaded from memory to a 128-bit XMM register. 
This is usually the case before invoking \texttt{AESKEYGENASSIST}, which is used to derive the AES round keys. 
The round-key derivation is entirely done in hardware using the XMM registers. 
Hence, there is no memory load required for the derivation of the 11 round keys used in AES-128. 
Thus, when the key is loaded from memory before the round-key derivation starts is the point where we can mount \AttackName to leak the value of the key. 
For OpenSSL (v3.0.0), this is in the function \texttt{aesni\_set\_encrypt\_key} which is called by \texttt{EVP\_EncryptInit\_ex}. 
Note that instead of leaking the key, we can also leak the round keys loaded in the encryption process. 
However, to attack the round keys, an attacker needs to leak (and distinguish) more different values, making the attack more complex. 

When leaking the key using \AttackName, we have first to detect which load corresponds to the key. 
Moreover, as we can only leak one byte at a time, we also have to combine the leaked bytes to the full AES-128 key correctly. 

\paragrabf{Side-Channel Synchronization.}
For the attack, we assume a shared library implementing the AES encryption which can be used by both the attacker and the victim, \eg OpenSSL. 
Even though OpenSSL (v3.0.0) has a side-channel resistant AES-NI implementation, we can still rely on classical memory-based side-channel attacks to monitor the control flow. 
For example, using \FlushReload, we can detect when a specific part of the code is executed~\cite{Gruss2015Template,Garcia2017constant}. 
While this does not leak any secrets, it acts as a synchronization primitive for \AttackName. 

We constantly monitor a cache line of the code which is executed right before the key is loaded from memory. 
In OpenSSL (v3.0.0), this is the second cache line of \texttt{aesni\_set\_encrypt\_key}, \ie \SI{64}{\byte} after the start of the function. 
Similarly to Schwarz~\etal\cite{Schwarz2018DF}, we leverage the cache state of the cache line as a trigger for the actual attack. 
Only if we detect a cache hit on the monitored cache line, we start leaking values using \AttackName. 
Hence, we already filter out most bytes not related to the AES key. 

Note that if there is no cache line before the load which can be used as a trigger, we can still use a nearby cache line (\ie a cache line after the load) as a filter. 
In a parallel thread, we collect the timestamps of cache hits in the nearby cache line. 
If we also save the time stamps of the values leaked using \AttackName, in an offline post-processing step we can filter out values which were leaked at a different instruction-pointer location. 

To further reduce unrelated loads, it is also possible to slow down the victim using performance-degradation techniques such as flushing the code~\cite{Allan2016degrade,Garcia2017constant}. 
For OpenSSL, we used performance degradation on the code directly following the load of the key. 

\paragrabf{Domino Attack.}

\begin{figure}
 \centering
 \resizebox{\hsize}{!}{
\begin{tikzpicture}[yscale=0.65]

\draw[fill=green!10,ultra thick] (4,0.4) rectangle +(8,2);
\node at (8,2.9) {\Huge $\text{{\huge(4,4)-}domino}_{n,n+1}$ (\texttt{0x21})};

\draw[fill=none,densely dashed,ultra thick] (1,-0.4) rectangle +(8,2);
\node at (5,-1.25) {\Huge $\text{{\huge(7,1)-}domino}_{n,n+1}$ (\texttt{0xA4})};

\draw (0,0) rectangle +(8,2);
\node[right] at (0.2,1) {\Huge\textbf{1}};
\node[right] at (1.2,1) {\Huge\textbf{1}};
\node[right] at (2.2,1) {\Huge\textbf{0}};
\node[right] at (3.2,1) {\Huge\textbf{1}};
\node[right] at (4.2,1) {\Huge\textbf{0}};
\node[right] at (5.2,1) {\Huge\textbf{0}};
\node[right] at (6.2,1) {\Huge\textbf{1}};
\node[right] at (7.2,1) {\Huge\textbf{0}};

\node at (2,2.5) {\Huge $\text{key}_n$ (\texttt{0xD2})};

\draw (8,0) rectangle +(8,2);
\node[right] at (8.2,1) {\Huge\textbf{0}};
\node[right] at (9.2,1) {\Huge\textbf{0}};
\node[right] at (10.2,1) {\Huge\textbf{0}};
\node[right] at (11.2,1) {\Huge\textbf{1}};
\node[right] at (12.2,1) {\Huge\textbf{1}};
\node[right] at (13.2,1) {\Huge\textbf{1}};
\node[right] at (14.2,1) {\Huge\textbf{0}};
\node[right] at (15.2,1) {\Huge\textbf{0}};

\node at (14,2.5) {\Huge $\text{key}_{n+1}$ (\texttt{0x1C})};

\end{tikzpicture}
}
 
 \caption{Additionally leaking domino bytes comprised of bits of different AES-key bytes to filter out unrelated loads.}
 \label{fig:domino}
\end{figure}
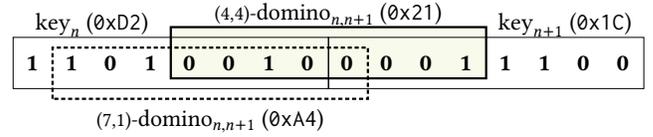

Inevitably, even when synchronizing \AttackName by using a cache-based trigger, we also leak values not related to the key. 
Moreover, for practical reasons, the size of the Flush+Reload covert channel is limited, and we can only transmit a single key byte from the transient domain at a time. 
Hence, we have a probability distribution for every byte in the AES key. 
As the bytes in the AES key are independent of each other, we can only assume that the byte with the highest probability is the correct key byte. 
Thus, if there is a key byte suffering from noise from unrelated loads, we may assume that the noise is the correct key byte, which leads to a wrong key. 

Therefore, we propose the \textit{Domino attack}, an innovative transient error detection technique for reducing noise when leaking multi-byte loads. 
In addition to leaking every single key byte, we also transmit a specially crafted \textit{domino byte} composed by combining bits from two adjacent key bytes.
Note that creating such a domino byte is possible, as the transient domain has access to the full AES key and can use it for arbitrary computations (\cf \Cref{sec:attack-covert-channel}). 
\Cref{fig:domino} illustrates the idea of the Domino attack. 
In this case, we leak (4,4) domino bytes consisting of 4 bits of two adjacent key bytes respectively. 
By combining the lower nibble of one key byte with the higher nibble of the next key byte, we transmit a domino byte which encodes partial information of two key bytes. 
Hence, in a post-processing step, we combine the probability distribution of two adjacent key bytes with the probability distribution of the domino byte to select the two adjacent key bytes with the highest combined probability. 
Note that the selection of bits can be adapted to the noise which can be measured before leaking the key, \eg multiple (7,1) domino bytes can be leaked that are shifted by only a single bit. 

\paragrabf{Results.}
We evaluated the attack in a cross-user-space attack (\cf \Cref{sec:attack-scenarios}). 
We always ran the attack until the correct key was recovered, \ie until the key with the highest probability is the correct key. 
In a practical attack, the number of attacks can even be reduced, as typically it is easy to verify whether a key candidate is correct. 
Thus, an attacker can simply test all key candidates with a probability over a certain threshold and does not have to wait until the highest probability corresponds to the correct key. 

On average, we recovered the entire AES-128 key of the victim in under \SI{10}{\second} using the cache-based trigger and the Domino attack. 
During this time, the key was loaded approximately \SI{10000} times by the victim. 

\subsection{SGX Sealing Key Extraction}\label{sec:attack-seal}
In this section, we show that privileged SGX attackers can drastically improve \AttackName's temporal resolution and bridge from incidental data sampling in the time domain to the targeted reconstruction of arbitrary enclave secrets (\cf \cref{fig:triangle}).
We first explain how state-of-the-art enclave execution control and transient post-processing techniques can be leveraged to reliably leak register values at any point during an enclave invocation.
Then we demonstrate the impact of this attack by recovering a full 128-bit SGX sealing key, as used by Intel's trusted provision and quoting enclaves to decrypt the long-term EPID private attestation key.

\paragrabf{Leaking Enclave Registers.}

We consider Intel SGX root attackers that co-locate with a victim enclave on the same physical CPU.
As a system attacker, we can increase \AttackName's temporal resolution by leveraging previous research results exploiting page faults~\cite{Xu2015controlled,Vanbulck2017pagetable} or interrupts~\cite{Vanbulck2018nemesis,Moghimi2017cachezoom} to regulate the victim enclave's execution.
We use the SGX-Step~\cite{VanBulck2017sgx} framework to precisely single-step the victim enclave one instruction at a time, allowing the attacker to reach a code part where sensitive information is stored in CPU registers.
At such a point, we switch to unlimited zero-stepping~\cite{Vanbulck2018} by either setting the system timer interrupt to a very short interval or revoking code page execute permissions before resuming the victim enclave.
This technique provides \AttackName attackers with a primitive to repeatedly force-reload CPU registers from the interrupted enclave's SSA frame (\cf \cref{sec:sgx}).
Our experiments show that even though execution of the enclave instruction never completes, any direct operands plus SSA register file contents are loaded from memory each time.
Importantly, since the enclave does not make progress, we can perform unlimited \AttackName attack attempts to reconstruct CPU register values from these implicit SSA memory accesses.

We further reduce noise from unrelated non-enclave loads on the victim CPU by opting for timer-based zero-stepping with a user space interrupt handler~\cite{Vanbulck2018nemesis} to avoid repeatedly invoking the operating system.
Furthermore, we found that executing the \AttackName attack code in a separate address space avoids unnecessarily slowing down the spy through implicit TLB invalidations on enclave entry/exit~\cite{sgxdeveloperref}.

Note that the SSA frame spans multiple cache lines. 
With \AttackName, we do not have explicit address-based control over which cache line is being leaked. 
Hence, leaked data might come from different saved registers that are at the same offset within a cache line. 
To filter out such noisy observations, we use the Domino transient error detection technique introduced in \cref{sec:attack-aes}.
Specifically, we implemented a \enquote{sliding window} that transmits 7 different domino bytes for each candidate key byte, stuffed with increasing bits from the next adjacent key byte candidate.
Any noisy observations that do not match the overlap can now efficiently be filtered out.

\paragrabf{Attack on \texttt{sgx\_get\_key}.}
The Intel SGX design includes a secure key derivation facility through the \texttt{egetkey} instruction (\cf \Cref{sec:background-extensions}).
Enclaves execute this instruction to query a 128-bit cryptographic key from the hardware, based on the calling enclave's code layout or developer identity.
This is the underlying primitive used by Intel's trusted prebuilt quoting enclave to securely unseal a long-term private attestation key from persistent storage~\cite{Costan2016,Vanbulck2018}.

The official Intel SGX SDK~\cite{sgxdeveloperref} offers a convenient \texttt{sgx\_get\_key} wrapper procedure that first executes \texttt{egetkey} with the necessary parameters, and eventually copies the retrieved key into a provided buffer.
We reverse engineered the proprietary \texttt{intel\_fast\_memcpy} function and found that in this case, the key is copied using two 128-bit moves to/from the \texttt{xmm0} SSE register.
We revert to zero-stepping on the last instruction of the \texttt{memcpy} invocation.
At this point, the attacker-induced zero-step enclave resumptions will repeatedly reload a.o., the \texttt{xmm0} register containing the 128-bit key from the memory hierarchy.

\paragrabf{Results.}
We evaluated the attack on a Kaby Lake i7-7700 CPU with an up-to-date Foreshadow-patched microcode revision 0x8e. 

In the first experiment, we implemented a benchmark enclave that uses \texttt{sgx\_get\_key} to generate a new report key with different random key IDs.
We performed 100 key-recovery experiments on \texttt{sgx\_get\_key} with different random keys. 
Our results show that \SI{30}{\percent} of the times the full 128-bit key is among the key candidates with average remaining key space entropy of 8.8 bits.
Among these cases, \SI{3}{\percent} of the times the exact full key has been recovered. 
In the other \SI{70}{\percent} of the cases where the full key is not among the key candidates, \SI{31}{\percent} of the times, we have partial key bytes among the recovered key candidates.
The average correct key bytes are 10 out of 16 bytes with the remaining global entropy of 13.59 bits.  
In the remaining \SI{39}{\percent} of the times where the correct key is not among the key candidates, our attack which uses the Domino technique with a sliding window did not reveal any candidates, which means an attacker can simply repeat the attack in such cases.
Also in cases, where some of the key bytes are part of the candidates, most of failed key bytes resides in the first few bytes of the key.
The reason for this behavior is that the explained Domino attack will have a stronger effect on key bytes in the middle that are surrounded by more key bytes.

In the second experiment, we perform an attack on Intel's trusted quoting enclave.
The quoting enclave performs a call to \texttt{sgx\_get\_key} to derive the sealing key which is used to decrypt 
the EPID provisioning blob.
We executed the attack on a quoting enclave that is signed with debug keys, so we can use it as a ground truth to easily verify that we have recovered the correct sealing key.
We executed the attack multiple times on our setup, and we managed to recover the correct 128-bit sealing key after multiple executions of the attack and checking the candidates against each other.
The recovered sealing key matches the correct key, and can indeed successfully decrypt the EPID blob for our debug signed quoting enclave. 
While we did not yet reproduce this attack to recover the sealing key from the official quoting enclave image signed by Intel, we believe that this experimental evaluation showcased all the required primitives to break Intel SGX's remote attestation guarantees, as demonstrated before by Foreshadow~\cite{Vanbulck2018}.

\subsection{Cross-VM Covert Channel}\label{sec:attack-covert-channel}

To evaluate the performance of \AttackName, we implement a covert channel which can be used for all attack scenarios described in \Cref{sec:attack-scenarios}. 
However, in this section, we focus on the cross-VM covert channel. 
While covert channels are possible for Intel SGX, the kernel, and the hypervisor, these are somewhat artificial scenarios. 
Moreover, there are various covert channels available to user-space applications for stealthy inter-process communication~\cite{Ge2016,Maurice2017Hello}. 

For VMs, however, there are not many known covert channels which can be used between two VMs. 
So far, all cross-VM covert channels either relied on \PrimeProbe~\cite{Ristenpart2009,Xu2011,Liu2015,Maurice2015C5,Maurice2017Hello}, DRAMA~\cite{Pessl2016}, or bus locking~\cite{Wu2012}. 
We show that \AttackName can be used as a fast and reliable covert channel between VMs scheduled on the same physical core. 

\paragrabf{Sender.}
For the fastest result, the sender repeatedly loads the value to be transmitted from the L1 cache into a register. 
By not only loading the value from one memory address but instead from multiple memory addresses, the sender ensures that potentially multiple fill-buffer entries are used. 
In addition, this also thwarts an optimization of Intel CPUs which combines multiple loads from the same cache line to a single load~\cite{Abramson1996}. 

On a CPU supporting AVX2, the sender can encode up to 256 bits per load (\eg using the \texttt{VMOVAPS} load).

\paragrabf{Receiver.}
The receiver mounts \AttackName to leak the values loaded by the sender. 
However, as the receiver leaks the loads only in the transient domain, the leaked value have to be transferred into the architectural domain. 
We encode the leaked values into the cache and recover them using \FlushReload. 
When encoding values in the cache, we require at least 2 cache lines, \ie \SI{128}{\byte}, per bit to prevent the adjacent-cache-line prefetcher from interfering with the encoding. 
In practice, we require one physical page, \ie \SI{4}{\kilo\byte}, per possible value to prevent interference of the prefetcher. 
To reduce the recover bottleneck, we transfer single bytes from the transient to the architectural domain which already requires 256 runs of \FlushReload. 

As a result, our proof-of-concept limits the transmission of actual data to a single byte per leaked load. 
However, we can use the remaining bits in the load to ensure that the channel is free of errors. 

\paragrabf{Transient Error Detection.}

\begin{figure}
 \centering
 \resizebox{\hsize}{!}{
    \begin{tikzpicture}[yscale=.5]

\draw (0,0) rectangle +(8,1);
\draw[fill=black!10] (0,0) rectangle +(2,1) node[midway] {\texttt{0xFF}};
\draw (2,0) rectangle +(2,1) node[midway] {\texttt{SEQ}};
\draw (4,0) rectangle +(2,1) node[midway] {$\overline{\mathtt{DATA}}$};
\draw (6,0) rectangle +(2,1) node[midway] {\texttt{DATA}};

\node at (7.9,1.25) {\scriptsize 0};
\node at (5.9,1.25) {\scriptsize 7};
\node at (3.9,1.25) {\scriptsize 15};
\node at (1.9,1.25) {\scriptsize 23};

\end{tikzpicture}
 }
 \caption{The packet format used in the covert channel. 
 Every 32-bit packet consists of 8 data bits, 8-bit checksum (two's complement), 8-bit sequence number, and a constant prefix.} 
 \label{fig:packet}
\end{figure}
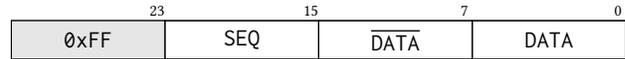

The transmission of the data between sender and receiver is free of any noise. 
However, the receiver does not only recover values from the sender, but also other loads from the current and sibling logical core. 
Hence, to get rid of this noise, we encode the data as shown in \Cref{fig:packet}. 
This allows the receiver to filter out data not originating from the sender. 

Although we cannot transfer the entire packet into the architectural domain, we can compute on the packet in the transient domain. 
Thus, we run the error detection in the transient domain, and only transmit valid packets to the architectural domain. 

The challenge to run the error detection in the transient domain is that the number of instructions is limited, and not all instructions can be used. 
For reliable results, we cannot use instructions which speculate on either control or data flow. 
Hence, the error-detection code has to be as short as possible and branch free. 

Our packet structure allows for extremely efficient error detection. 
We encode the data in the first byte and the two's complement of the data in the second byte as a checksum. 
To detect errors, we \texttt{XOR} the value of the first byte (\ie the data) onto the second byte (\ie the two's complement of the data). 
If both values are received correctly, the \texttt{XOR} ensures that the bits 8 to 15 of the packet are zero. 
Thus, for a correct packet, the least-significant 16 bits of the packet represent a value between 0 and 255, and for a wrong packet, these bits represent a value which is larger than 255. 
We use these resulting 16-bit value as an index into our oracle array, \ie an array consisting of 256 pages. 
Therefore, any value which is not a correct byte is out of bounds and has thus no effect on the cache state of the array. 
A correct byte is also a valid index into the oracle array and ensures that the first cache line of the corresponding page is cached. 
Finally, by applying a cache-based side-channel attack, such as \FlushReload, we can recover the byte from the cache state of the oracle array~\cite{Lipp2018meltdown,Kocher2019spectre}. 

The error detection in the transient domain has the advantage that we do not require computation time in the architectural domain. 
Instead of waiting for the exception to become architecturally visible by doing nothing, we already use this time to perform the required computation. 
An additional advantage is that while we are still in the transient domain, we can work on noise-free data. 
Thus, we do not require complex error correction after receiving the data~\cite{Maurice2017Hello}. 

In addition to the error detection, we also encode a sequence number into the packet. 
The sequence number allows ordering the received packets. 
It can be recovered using the same method as the data value, \eg using an oracle array and a cache-based side-channel attack.

\paragrabf{Results.}

We evaluate the covert channel both in a lab environment as well as in a public cloud. 
In the lab environment, we used 2 virtual machines running inside QEMU KVM on an i7-8650U. 
For the cloud scenario\footnote{The cloud provider asked us not to disclose its name at this point.}, we used 2 co-located virtual machines running CentOS 7.6.1810 with a Linux kernel version of 3.10.0-957 on a Xeon E5-2670 CPU.

Both on the cloud, as well as on our lab machine, we achieved an error-free transmission. 
On our lab machine, we observed transmission rates of up to \SI{26.8}{\kilo\bit/\second}.
As TSX was not available in the cloud scenario, we achieved a transmission rate of \SI{1.99}{\kilo\bit/\second} ($\sigma_{\bar{x}}=\SI{2.5}{\percent}$, $n=1000$) with \VariantOne and signal handling.

\subsection{Browsing-Behavior Monitoring}\label{sec:attack-browser}

\AttackName is also well suited for detecting specific byte sequences within loaded data. 
We demonstrate an attack for which we leverage \AttackName to fingerprint a web browser session. 
For this attack, we assume an unprivileged attacker running on one logical core and a web browser running on the sibling logical core. 
In this scenario, it is irrelevant whether the attacker and victim run on a native machine or whether they are in (different) virtual machines. 

We present two different attacks, a keyword detection attack which can fingerprint website content, and an URL recovery attack to monitor a victim's browsing behavior. 

\paragrabf{Keyword Detection.}
The keyword detection allows an attacker to gain information on the type of content the victim is consuming. 
For this attack, we constantly sample data using \AttackName and match leaked values against a list of pre-defined keywords. 

We leverage the fact that we have access to a full cache line and can do arbitrary computations in the transient domain (\cf \Cref{sec:attack-covert-channel}).
As a result of the computation, we only have to externalize a small integer indicating which keyword has matched via a cache side channel.

One limitation is the length of the keyword list, as in the transient domain, only a limited number of memory accesses are possible before the transient execution aborts. 
The most reliable solution is to store the keyword list entirely in CPU registers. 
Hence, the length of the keyword list is limited by the available registers. 
Moreover, the length is also limited by the amount of code that is transiently executed to compare leaked values to the keyword list.

\paragrabf{URL Recovery.}
In the second attack, we recover accessed websites from browser sessions without prior selection of interesting keywords. 
We take a more indirect approach that relies on modern websites performing many individual HTTP requests to the same domain, \eg to load additional resources such as scripts and images. 

In the transient domain, we again sample data using \AttackName.
While still in the transient domain, we detect the substring ``\texttt{www.}'' inside the leaked data. 
When we discover a match, we leak the character following ``\texttt{www.}'' to the architectural domain using a cache side channel. 
This already results in a set of first characters of domain names which we refer to as the candidate set. 

In the next iteration, for every domain in the candidate set, we take the last four leaked characters (\eg ``\texttt{ww.X}'').
We use this string in the transient domain to filter leaked values, similar to the ``\texttt{www.}'' substring in the first iteration. 
If a match is found, we leak the next character. 
We can repeat these steps until we see a string ending with a top-level domain.

Note that this attack is not limited to URLs. 
Potentially all data which follows a predictable pattern, such as session cookies or credit-card numbers, can be leaked with this variant. 

\paragrabf{Results.}

We evaluated both attacks running an unmodified Firefox browser version 66.0.2 on the same physical core as the attacker.
Our proof-of-concept implementation of the keyword-checking attack can check four up to 8-byte long keywords.
Due to excessive precomputations of browsers when entering an URL, a keyword is sometimes already matched during the autocompletion of the URL.
For highly dynamic websites, such as \emph{nytimes.com}, keywords reliably match on the first access of the website.
Accessing mostly static websites, such as \emph{gnupg.org}, have a \SI{60}{\percent} probability of matching a keyword in this setup.
We observed false positives after the first website access when continuing to use the browser.
We hypothesize that memory locations containing the keywords get re-used and may thus leak at a later time again.

For the URL recovery attack, we simulated user behavior by accessing popular websites and refreshing them in a defined time interval.
We counted the number of refreshes necessary until we recovered the entire URL including top level domain.
For each website, the experiment was repeated 100 times.

\begin{table}
\caption{Number of accesses required to recover a website name. The experiment was repeated 100 times per website.}
\label{tab:url_recover}
\begin{tabular}{r|rrr}
\textbf{Website} & \textbf{Minimal} & \textbf{Average} & \textbf{Maximum} \\\hline
nytimes.com  &  1 &   1 &   3 \\
facebook.com &  1 &   2 &   4 \\
kernel.org   &  2 &   6 &  13 \\
gnupg.org    &  2 &  10 &  34 \\
\end{tabular}
\end{table}

The actual number of refreshes needed depends on the nature of the website that is visited.
If it is a highly dynamic page, such as \emph{facebook.com} or \emph{nytimes.com}, a small
number of reloads is sufficient to recover the entire name.
For static pages, such as \emph{gnupg.org} or \emph{kernel.org}, the number of reloads necessary
increases by a factor of 10, approximately. See \Cref{tab:url_recover} for a detailed overview
of required reloads.
 
\subsection{Targeted Data Leakage}\label{sec:attack-data}

Inherently, \AttackName is a 1-dimensional side channel, \ie the leakage is only controlled by the time. 
Hence, leakage cannot be steered using specific addresses as is the case, \eg for Meltdown~\cite{Lipp2018meltdown}. 
While this data sampling is still sufficient for several real-world attacks, it is still a limiting factor for general attacks. 

In this section, we show how \AttackName can be combined with \textit{prefetch gadgets}~\cite{Canella2019} for targeted data leakage. 

\begin{listing}[t]
 \begin{lstlisting}[style=customc]
if (x < array_len) {
    y = array[x];
} \end{lstlisting}
\caption{A simple prefetch gadget relying on Spectre-PHT~\cite{Kocher2019spectre}.
By mistraining the branch, this gadget loads an arbitrary out-of-bounds value for targeted leakage.}
\label{lst:prefetch-gadget}
\end{listing}

\paragrabf{Speculative Data Leakage.}
\Cref{lst:prefetch-gadget} illustrates such a gadget. 
It is a common pattern in software for accessing an element of an array~\cite{Canella2019}. 
First, the code checks whether the index lies within the bounds of the array. 
Only if this is the case, the element is accessed, \ie loaded. 
While it is evident that for a user-controlled index the corresponding array element can be loaded, such a gadget is even more powerful. 

On a CPU vulnerable to Spectre, an attacker can mistrain the branch predictor, \eg by providing several valid values for the array index.
Then, by providing an out-of-bounds index, the branch is misspeculated and speculatively accesses an out-of-bounds value.
Alternatively, the attacker can alternate between valid and out-of-bounds indices randomly to achieve a high percentage of mispredictions without any prior branch predictor mistraining.

\AttackName cannot only leak architecturally accessed data but also speculatively accessed data. 
Hence, \AttackName can even see the value of loads which are never architecturally visible. 
Such loads include, among others, speculative memory loads and prefetches. 
Thus, any Spectre gadget which is not hardened, \eg using a memory fence~\cite{IntelSpecAnalysis,AMDSpecAnalysis,ARMSpecAnalysis,Canella2019} or a mask~\cite{Carruth2018Hardening,Canella2019}, can be used to specify data to leak. 

Moreover, \AttackName does not require classic Spectre gadgets containing an indirect array access~\cite{Kocher2019spectre}. 
A simple out-of-bounds access (\cf \Cref{lst:prefetch-gadget}) is sufficient. 
While such gadgets have been demonstrated for breaking KASLR~\cite{Schwarz2018netspectre}, they were considered as relatively harmless as they do not leak data~\cite{Canella2019}.
Hence, most approaches for finding gadgets do not consider such gadgets~\cite{Wang2017oo7,Guarnieri2018spectector}. 
In the Linux kernel, however, such gadgets are also patched if they are discovered, mainly as they can be used together with the Foreshadow vulnerability to leak arbitrary kernel memory~\cite{Corbet2018smatch,Stecklina2019L1TF}.
So far, 172 such gadgets have been fixed in kernel 5.0~\cite{Canella2019}. 
With \AttackName, we show that such gadgets are indeed powerful and have to be patched as well.

\paragrabf{Potential Incompleteness of Countermeasures.}
Mainly, there are 2 methods to prevent exploitation of Spectre-PHT: memory fences after branches~\cite{IntelSpecAnalysis,AMDSpecAnalysis,ARMSpecAnalysis,Canella2019}, or constraining the index to a valid range using a bitmask~\cite{Carruth2018Hardening,Canella2019}. 
The variant using fences is implemented in the Microsoft compiler~\cite{Kocher2018mitigations,Kocher2019spectre}, whereas the variant using bitmasks is implemented in GCC~\cite{LWN_GCC_SLH} and LLVM~\cite{Carruth2018Hardening}, and also used in the Linux kernel~\cite{LWN_GCC_SLH}. 

Both methods prevent exploitation of Spectre-PHT~\cite{Canella2019}, as the misspeculation cannot load any data. 
Hence, this is also effective against \AttackName, as fixed gadgets cannot be exploited to load arbitrary values. 

However, even with these countermeasures in place, there is a remaining leakage which can be exploited using \AttackName. 
When architecturally loading an in-bounds value, \AttackName can leak up to 64 bytes of the load. 
Hence, with \AttackName, there is a potential leakage of up to 63 bytes which are out of bounds if the last in-bounds value is at the beginning of a cache line or the base of the array is at the end of a cache line. 

\paragrabf{Data Leakage.}

To demonstrate the feasibility of prefetch gadgets for targeted data leakage, we leverage an artificial prefetch gadget as given in \Cref{lst:prefetch-gadget}. 
For our evaluation, we used such a gadget in the system-call path of the Linux kernel 5.0.7.
We execute \AttackName on one logical core and on the other we execute system calls that switch between out-of-bounds and in-bounds array indices to achieve a high frequency of mispredictions in the gadget.

This approach yields leaked values with a large noise component from unrelated loads.
We repeat this setup without trying to generate mispredictions to generate a baseline of noise values.
We generate frequency distributions for both runs and subtract the noise frequency from the misprediction run.
We then choose the byte value that was seen most frequently.

With this crude statistical method, we can recover kernel memory at one byte per \SI{10}{\second} with \SI{38}{\percent} accuracy.
Probing bytes for \SI{20}{\second} improves the accuracy to \SI{46}{\percent}.

As with Meltdown~\cite{Lipp2018meltdown}, common byte values such as \texttt{0x00} and \texttt{0xFF} occur too often and have to be removed from the leaked data for the recovery to work.
Our approach is thus blind to these values.

The speed and accuracy can be improved if there is a priori knowledge of the target data. 
For example, a 7-bit ASCII string can be leaked with a probing time of \SI{10}{\second} per byte with \SI{72}{\percent} accuracy.

\section{Countermeasures}\label{sec:countermeasures}

As \AttackName leaks loaded values across logical cores, a straight-forward mitigation is disabling the use of hyperthreading. 
Hyperthreading improves performance for certain workloads by \SI{30}{\percent} to \SI{40}{\percent}~\cite{bulpin2004multiprogramming,Phoronix2018HT}, and as such disabling it may incur an unacceptable performance impact.

\paragrabf{Co-Scheduling.}
Depending on the workload, a more efficient mitigation is the use of co-scheduling~\cite{Ousterhout1982}. 
Co-scheduling can be configured to prevent the execution of code from different protection domains on a hyperthread pair.
Current topology-aware co-scheduling algorithms~\cite{Schoenherr2013} are not concerned with preventing kernel code from running concurrently with user-space code. 
With such a scheduling strategy, leaks between user processes can be prevented but leaks between kernel and user space cannot. 
To prevent leakage between kernel and user space, the kernel must additionally ensure that kernel entries on one logical core force the sibling logical core into the kernel as well. 
This discussion applies in an analogous way to hypervisors and virtual machines.

\paragrabf{Flushing Buffers.}
We have demonstrated that \AttackName also works across protection boundaries on a single logical core. 
Hence, disabling hyperthreading or co-scheduling are not fully effective as mitigation. 
We have not found an instruction sequence that reliably prevents leakage across protection boundaries.
Even flushing the entire L1 data cache (using \texttt{MSR\_IA32\_FLUSH\_CMD}) and issuing as many dummy loads as there are fill-buffer entries (``load stuffing'') is not sufficient. 
There is still remaining leakage, which we assume is caused by the replacement policy of the line-fill buffer. 
Hence, to fully mitigate the leakage, we require a microcode update which provides a method to flush the line-fill buffer. 

\paragrabf{Selective Feature Deactivation.}
Weaker countermeasures target individual building blocks (\cf \Cref{sec:building-blocks}). 
The operating system kernel can make sure always to set the accessed and dirty bits in page tables to impair \VariantTwo. 
Unfortunately, \VariantOne is always possible, if the attacker can identify an alias mapping of any accessible user page in the kernel. 
This is especially true if the attacker is running in or can create a virtual machine.
Hence, we also recommend disabling VT-x on systems that do not need to run virtual machines.

\paragrabf{Removing Prefetch Gadgets.}
To prevent targeted data leakage, prefetch gadgets need to be neutralized, \eg using \textit{array\_index\_nospec} in the Linux kernel. 
This function clamps array indices into valid values and prevents arbitrary virtual memory to be prefetched. 
Placing these functions is currently a manual task and due to the incomplete documentation of how Intel CPUs prefetch data, these mitigations cannot be complete. 
Note that Spectre mitigations using \texttt{lfence} instructions might also be incomplete against \AttackName.

Another way to prevent prefetch gadgets from reaching sensitive data is to prevent this data from being mapped in the address space of the prefetch gadget. 
Exclusive Page-Frame Ownership~\cite{Kemerlis2014} (XPFO) partially achieves this for the Linux kernel's mapping of physical memory.

Prefetch gadgets can also be neutralized using Speculative Load Hardening~\cite{Carruth2018Hardening} (SLH). 
SLH prevents speculative execution by introducing artificial data dependencies via a compiler pass. 
SLH incurs a performance overhead of \SI{10}{\percent} to \SI{50}{\percent} for typical applications.
To the best of our knowledge, its overhead for kernel or hypervisor code has not been studied yet.

\paragrabf{Instruction Filtering.}
The above discussion mostly focusses on attacks across process or virtual-machine boundaries. 
For attacks inside of a single process (\eg JavaScript sandbox), the sandbox implementation must make sure that the requirements for mounting \AttackName are not met. 
One example is to prevent the generation and execution of the \texttt{clflush} instructions, which so far is a crucial part of the attack.

\paragrabf{Secret Sharing.}
On the software side, we can also rely on secret sharing techniques used to protect against physical side-channel attacks~\cite{Shamir1979secretsharing}.
We can ensure that a secret is never directly loaded from memory but instead only combined in registers before being used. 
As a consequence, observing the data of a load does not reveal the secret. 
For a successful attack, an attacker has to leak all shares of the secret. This mitigation is, of course, incomplete if register values are written to and subsequently loaded from memory as part of context switching.

\section{Conclusion}\label{sec:conclusion}

With \AttackName, we showed a novel Meltdown-type attack targeting the processor's fill-buffer logic.
\AttackName enables an attacker to leak recently loaded values used by the current or sibling logical CPU.
We show that \AttackName allows leaking across user-space processes, CPU protection rings, virtual machines, and SGX enclaves.
We demonstrated the immense attack potential by monitoring browser behaviour, extracting AES keys, establishing cross-VM covert channels or recovering SGX sealing keys.
Finally, we conclude that disabling hyperthreading is the only possible workaround to mitigate \AttackName on current processors. 

\section*{Acknowledgments}
We thank Werner Haas (Cyberus Technology), Claudio Canella (Graz University of Technology), Jon Masters (Red Hat), Alex Ionescu (CrowdStrike), and Martin Schwarzl (Graz University of Technology).
The research presented in this paper was partially supported by the Research Fund KU Leuven.
Jo Van Bulck is supported by a grant of the Research Foundation -- Flanders (FWO).
The project was supported by the European Research Council (ERC) under the European Union's Horizon 2020 research and innovation programme (grant agreement No 681402).
It was also supported by the Austrian Research Promotion Agency (FFG) via the K-project DeSSnet, which is funded in the context of COMET – Competence Centers for Excellent Technologies by BMVIT, BMWFW, Styria and Carinthia.
Additional funding was provided by a generous gift from Intel.
Any opinions, findings, and conclusions or recommendations expressed in this paper are those of the authors and do not necessarily reflect the views of the funding parties.

{\footnotesize \bibliographystyle{acm-url}
\bibliography{main}}

\appendix

\section{Fill-buffer Size}\label{appendix:fill-buffer-size}
In this section, we analyze the size of the fill buffer in terms of fill-buffer entries usable per logical core. 
Intel describes the fill buffer as a ``competitively-shared resource during HT operation''~\cite{Intel_vol3}. 
Hence, with 10 fill-buffer entries (Sandy Bridge and newer microarchitectures)~\cite{Intel_vol3}, we expect that when hyperthreading is enabled, every logical core can use up to 10 entries. 

Our experimental setup measures the time it takes to execute $n$ stores to DRAM, for $n = 1, \dots, 20$. 
We expect that the time increases linearly with the number of stores $n$ as long as there are unused fill-buffer entries. 
To ensure that the stores occupy the fill buffer, we leverage non-temporal stores which bypass the cache and directly go to DRAM. 
We repeated our experiments \SI{1000000} times, and we always measured the best case, \ie the minimum latency, to get rid of any noise. 

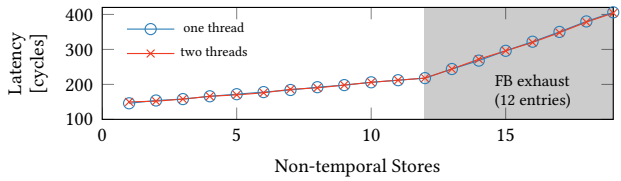
\begin{figure} 
\resizebox{\hsize}{!}{
 \begin{tikzpicture}
\begin{axis}[
style={font=\footnotesize},
xlabel={Non-temporal Stores},
ylabel={\parbox{1.5cm}{\centering Latency [cycles]}},
width=0.95\hsize,
xmin=0,
xmax=19,
ymin=100,
ymax=420,
scaled y ticks=false,
height=3cm,
legend pos=north west,
legend style={draw=none,fill=none}
]
\draw[fill=black, opacity=0.2,draw=none] (axis cs: 12,-2) rectangle (axis cs: 139,500);
\node at (axis cs: 16,180) {\parbox{2cm}{\centering \scriptsize{FB exhaust\\(12 entries)}}};

\node at (axis cs: 44,200) {\parbox{3cm}{\centering \scriptsize Latency increase\\(36 entries)}};
\draw[->,>=stealth] (axis cs: 36,200) to (axis cs: 36,112);


\addplot+[mark=o] table[x=stores,y=time,col sep=comma]{data/fill_buffer_one_ht.csv};

\addplot+[mark=x] table[x=stores,y=time,col sep=comma]{data/fill_buffer_two_ht.csv};

\legend{\tiny one thread,\tiny two threads}

\end{axis}
\end{tikzpicture} 
 }
 \caption{One logical core can leverage the entire fill buffer (12 entries). 
 If both logical cores execute stores, the fill buffer is competitively shared, leading to an increased latency for both logical cores.}
 \label{fig:fbsize-split} 
\end{figure}

\Cref{fig:fbsize-split} shows that both logical cores can indeed leverage the entire fill buffer. 
When running the experiment on one (isolated) logical core, while the other (isolated) logical core does nothing, we get a latency increase when executing more than 12 stores. 
When we run the experiment on both logical cores in parallel, the latency increase is still after 12 stores. 

\begin{figure} 
\resizebox{\hsize}{!}{
 \begin{tikzpicture}
\begin{axis}[
style={font=\footnotesize},
xlabel={Non-temporal Stores},
ylabel={\parbox{1.5cm}{\centering Latency [cycles]}},
width=0.95\hsize,
xmin=6,
xmax=15,
ymin=280,
ymax=540,
scaled y ticks=false,
height=3cm,
legend pos=north west,
legend style={draw=none,fill=none}
]
\draw[fill=black, opacity=0.2,draw=none] (axis cs: 73,-2) rectangle (axis cs: 139,300);
\node at (axis cs: 79,100) {\parbox{2cm}{\centering \scriptsize{FB exhaust\\(12 entries)}}};

\node at (axis cs: 10,450) {\parbox{3cm}{\centering \scriptsize Latency increase\\(10 entries)}};
\draw[->,>=stealth] (axis cs: 10,400) to (axis cs: 10,360);

\node at (axis cs: 13.9,350) {\parbox{3cm}{\centering \scriptsize Latency increase\\(12 entries)}};
\draw[->,>=stealth] (axis cs: 12.75,360) to (axis cs: 12,360);


\addplot+[mark=o] table[x=stores,y=time,col sep=comma]{data/fill_buffer_haswell.csv};
 
\addplot+[mark=x] table[x=stores,y=time,col sep=comma]{data/fill_buffer_skylake.csv};

\legend{\tiny Haswell,\tiny Skylake}

\end{axis}
\end{tikzpicture}
  
 }
 \caption{One pre-Skylake, we measure 10 fill-buffer entries, matching Intel's documentation. On Skylake and newer, we measure 12 fill-buffer entries.}
 \label{fig:fbsize-skylake} 
\end{figure}
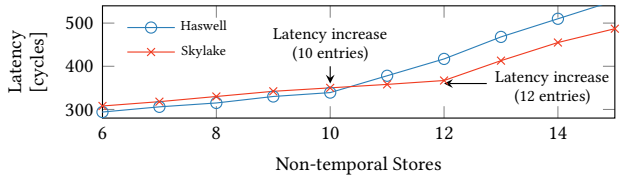

Interestingly, the documented number of fill buffers does not match our experiments for Skylake and newer microarchitectures. 
While we measure 10 entries on pre-Skylake CPUs as it is documented, we measure 12 entries on Skylake and newer (\cf \Cref{fig:fbsize-skylake}). 

From our experiments we conclude that both logical cores can leverage the entire fill buffer
Therefore, every logical core can potentially use any entry in the fill buffer. 

\end{document}